\documentclass[a4paper,fleqn,usenatbib]{mnras}

\usepackage[T1]{fontenc}
\usepackage{ae,aecompl}

\usepackage{graphicx}	
\usepackage{epstopdf}
\usepackage{amsmath}	
\usepackage{amssymb}	
\usepackage{comment}

\def\HI{\hbox{H~$\scriptstyle\rm I$}}

\title[AGN and reionization]{ 
	Constraining the UV emissivity of AGN throughout cosmic time via X-ray surveys}   

\author[F. Ricci et al.]   {Federica Ricci$^{1}$\thanks{riccif@fis.uniroma3.it}, Stefano Marchesi$^{2,3,4}$, Francesco Shankar$^5$,  
	\newauthor Fabio La Franca$^{1}$, Francesca Civano$^{4,6}$\\ \\ 
	\\ \\ 
	$^1$Dipartimento di Matematica e Fisica, Universit\`a Roma Tre, via della Vasca Navale 84, I-00146, Roma, Italy;\\
	$^2$Dipartimento di Fisica e Astronomia, Universit\`a di Bologna,
	viale Berti Pichat 6/2, 40127 Bologna, Italy\\
	$^3$INAF--Osservatorio Astronomico di Bologna, via Ranzani 1, 40127 Bologna, Italy\\
	$^4$Harvard-Smithsonian Center for Astrophysics, 60 Garden
	Street, Cambridge, MA 02138, USA\\
	$^5$Department of Physics and Astronomy, University of Southampton,
	Highfield, SO17 1BJ, UK\\
	$^6$Yale Center for Astronomy and Astrophysics, 260 Whitney Avenue, New Haven, CT 06520, USA.
}

\date{Accepted XXX. Received YYY; in original form April 2016}

\pubyear{2016}

\begin{document}
\label{firstpage}
\pagerange{\pageref{firstpage}--\pageref{lastpage}}
\maketitle

\begin{abstract}
The cosmological process of hydrogen (\HI) reionization in the intergalactic
medium is thought to be driven by UV photons emitted by star-forming galaxies
and ionizing active galactic nuclei (AGN). 
The contribution of QSOs to \HI\ reionization at $z>4$ has been traditionally believed to be quite modest. 
However, this view has been recently challenged by new estimates of a higher faint-end UV luminosity function (LF).  
To set firmer constraints on the emissivity of AGN at $z<6$, 
we here make use of complete X--ray selected samples 
including deep \textit{Chandra} and new COSMOS data, capable to efficiently 
measure the 1 ryd comoving AGN emissivity up to $z\sim5-6$ 
and down to five magnitudes fainter than probed by current optical surveys, without any luminosity extrapolation. 
We find good agreement between the logN$\rm{_H}\lesssim21-22$ cm$^{-2}$ X--ray LF and 
the optically-selected QSO LF at all redshifts for $M_{1450}\leq -23$. 
The full range of the 
logN$\rm{_H}\lesssim21-22$ cm$^{-2}$ LF ($M_{1450} \leq -17$) was then used to quantify the contribution of
AGN to the photon budget critical value needed to keep the Universe ionized. 
We find that the contribution of ionizing AGN at $z = 6$ is as small
as 1\% - 7\%, and very unlikely to be greater than 30\%, thus excluding
an AGN-dominated reionization scenario. 
	

\end{abstract}




\section{Introduction}


The transition from the so called dark ages to an ionized Universe involves the cosmological transformation of neutral hydrogen (\HI), 
which mostly resides in the intergalactic medium (IGM), into an ionized state. 
Observations of distant active galactic nuclei (AGN) and gamma-ray bursts set the 
end of this process to $z\sim6$ \citep{Fan2002,Kawai2006,McGreer2015}, 
as confirmed by both theoretical calculations \citep{Madau1999,MiraldaEscude2000,Choudhury2009} 
and numerous observational astrophysical evidences \citep{McGreer2011,Pentericci2011,Planck2014}. 
All these studies broadly constrain the epoch of hydrogen reionization
at $6<z<12$ (with a peak probability at $z\sim10$ if instantaneous reionization is assumed). 
The sources of ionizing photons (with energy greater than 13.6 eV, i.e. $\lambda\le912$ \AA) 
are traditionally believed to be star-forming galaxies (SFGs) and quasars (QSOs), 
though, which is the dominant one, is still a matter of considerable debate 
\citep{Haiman1998,SchirberBullock2003,ShankarMathur2007,Robertson2010,Bouwens2012,Fontanot2012}.
The global details of the \HI\ reionization are still elusive and one of the major 
	puzzles is that, given the properties of observed galaxies and AGN, none of the 
	known astrophysical populations is able to account alone for the whole ionizing photon budget 
	required to complete reionization at $z>6$, 
	thus not excluding a relevant contribution of more exotic sources 
	\citep{Scott1991,Madau2004,Pierpaoli2004,VolonteriGnedin2009,Dopita2011}. 

The fraction of ionizing photons 
that freely escape each galaxy,
$f_{esc}$, is expected to be low, given that 
galaxies are characterized by observed soft spectra blueward of Ly$\alpha$
due to the presence of cold gas and dust, 
which absorb most of the Lyman continuum emission \citep[see, e.g. ][]{Haehnelt2001}. 
The $f_{esc}$ of SFGs is however still not highly constrained, 
even though the 
general idea is that is much lower 
\citep[i.e., $f_{esc}\sim$ 0.1-0.2 at $z\sim$3-4,][and $f_{esc}\sim$ 0.05 - 0.1 at $z<1$, \citealt{Bridge2010,Barger2013}]{Shapley2006,Vanzella2010},
than the $f_{esc}$ of QSOs \citep[for different results see, e.g.][]{Fontanot2014,DuncanConselice2015,Rutkowski2015}.
Indeed, because of their observed UV hard spectra, some AGN are supposed to have large 
$f_{esc}$,
possibly reaching unity in the most luminous QSOs \citep[see][for a significant direct detection 
at $z = 3.46$, and also the results of \citealt{Cristiani2016}]{Guaita2016}.

Although QSOs have high $\langle f_{esc} \rangle$, it is traditionally assumed
that they are not the main contributors to \HI\ reionization due
to the steadily decreasing number density of AGN at $z>3$ \citep[e.g. ][]{Masters2012}.
Recently multiwavelenght deep surveys at $z>3$ \citep{Glikman2011,Fiore2012,Giallongo2015} detected 
a larger number density of faint AGN at high redshifts, thus possibly implying a more substantial AGN contribution to \HI\ reionization
\citep[][but see \citealt{Weigel2015,Cappelluti2015,Georgakakis2015,HS2015}]{MH2015}.

However, strongly UV emitting QSOs, showing optical blue spectra and broad emission lines (i.e. type-1 AGN), 
are only one class of the entire population of AGN, which also includes the type-2 AGN, 
characterised by red optical continuum and narrow emission lines. 
According to the standard AGN unified model \citep[e.g.][]{Urry1995}, 
this kind of classification depends on the observer line-of-sight 
with respect to an obscuring material, i.e. a dusty and probably clumpy torus.
In a more realistic scenario, also a contribution to the obscuration 
from the interstellar material of the hosting galaxy is expected \citep[e.g.][]{Granato2004,Granato2006}.
Therefore all AGN are alike but 
one source can appear as a type-1 or a type-2 depending on orientation and/or host galaxy properties.
Although in this scenario type-2 AGN could appear as type-1 UV emitting AGN 
under different line-of-sights, they should not be taken into account in the 
derivation of the UV background as 
in any direction only 
type-1 AGN contribute to the UV background and, for isotropic arguments,
their fraction should be the same \citep[see also][]{Cowie2009}.
In this framework, concerning the entire AGN population, $f_{esc}$ accounts for the fraction of unobscured AGN, 
since UV photons emitted by more obscured 
objects are likely to be absorbed locally within the host galaxy and therefore do not contribute to the cosmic reionization \citep[see also][]{Georgakakis2015}.

In the last decade, hard X--ray surveys have allowed to 
select almost complete AGN samples (including both type-1 and type-2 objects). Thanks to these studies, the evolution 
of the whole AGN population has been derived up to $z\sim5$ by many authors, all achieving fairly
consistent results \citep[][]{Ueda2003,LaFranca2005,Brusa2009,Civano2011,Ueda2014,Kalfountzou2014,Vito2014,Miyaji2015,Georgakakis2015,Aird2015,Aird2015Nustar}.
The X--ray spectra of AGN show a wide range of absorbing column densities ($20<{ \log \rm N_H}<26$ cm$^{-2}$),
with optically-classified type-1 AGN (i.e. QSOs) believed to broadly correspond to those AGN with the lowest N$_{\rm H}$ distributions, 
tipically $\log$N$\rm _H<$21 cm$^{-2}$,
though the exact correlation between X--ray and optical classifications is still quite debated
\citep[][]{Lusso2013,Merloni2014}.
In this framework, the X--ray luminosity function (XLF) of the 
AGN with low column densities ($\log$N$_{\rm H}<21-22$ cm$^{-2}$)
could be potentially used as an unbiased proxy 
of the ionizing AGN population (i.e. QSOs), where $\langle f_{esc}\rangle \sim1$ is expected.
On the contrary, at larger column densities, $f_{esc}$ should sharply decrease down to zero.
The advantage of X--ray selection is that it is less biased 
toward line-of-sight obscuration, extinction and galaxy dilution, especially at high $z$ (where harder portion of the spectra 
are probed), assuring a better handle on the faint-end of the 
AGN luminosity function (LF) compared to UV/optically selected samples. 
Additionally, at low luminosities, the standard optical color-color QSO identification procedure becomes less reliable, because QSO emission is superseded by the hosting galaxy. Moreover, moving to high redshifts, stars can be misinterpreted as QSOs: consequently, low-luminosity optical surveys have so far produced disagreeing QSO LFs \citep[QLFs, see][]{Glikman2011,Ikeda2011,Masters2012}. 

In this work we make use of the latest results on the X--ray AGN number densities including deep \textit{Chandra} and COSMOS data \citep[e.g.][]{Ueda2014,Vito2014,Marchesi2016b}.
Our aim is to provide more stringent constraints on the AGN contribution to the \HI\ reionization. 
To achieve this, we investigate whether the low N$_{\rm H}$ XLF can be used as an unbiased proxy to 
derive robust estimates of the QSO ionizing emissivity \citep[for a similar approach at $3<z<5$ see][]{Georgakakis2015}.
We will study the AGN LF up to redshift $\sim6$ over a 
broad range of luminosities, $10^{42}<L_X<10^{46.5}$ erg s$^{-1}$, five magnitudes fainter than the UV/optically-selected LFs, 
thus providing more stringent constraints on the density of low-luminosity QSOs. 

The paper is organized as follows: in Sect. \ref{sec:data} 
we describe the UV/optical and X--ray AGN LFs used in our study. 
In Sect. \ref{sec:uvx} we compare the UV/optical
and the X--ray LFs in order to 
determine which 
subsample of the XLF
better describes the UV/optical QSO LF, where Sect. \ref{sec:faintUV} focuses 
on the UV LF faint end at $z>4$. 
Sect. \ref{sec:emi} describes 
the computation of the ionizing AGN emissivity.
In Sect. \ref{sec:dis} the discussion is presented while in Sect. 
\ref{sec:conclusion} there are the conclusions.
Unless otherwise stated, all quoted errors are at the 68\% ($1\sigma$)
confidence level. Throughout the paper we assume a standard cosmology with parameters 
$\rm H_0=70$ km s$^{-1}$Mpc$^{-1}$, $\Omega_m = 0.3$ and $\Omega_\Lambda = 0.7$.

\section{DATA}
\label{sec:data}
We start off by comparing a relevant number of complete UV/optically selected QSO
and X--ray selected AGN samples and LFs.  

\subsection{QSO UV Luminosity Functions}\label{sec:QLF}
\begin{figure*}
	\centering
	\includegraphics[width=1.1\columnwidth]{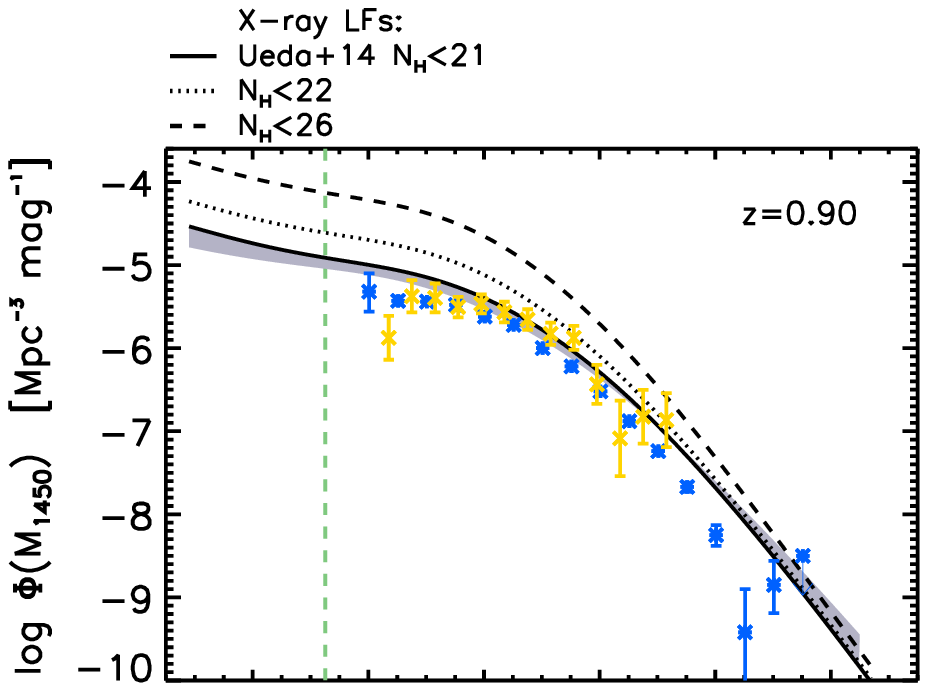}
	\hspace{-2.75cm}{\includegraphics[width=1.1\columnwidth]{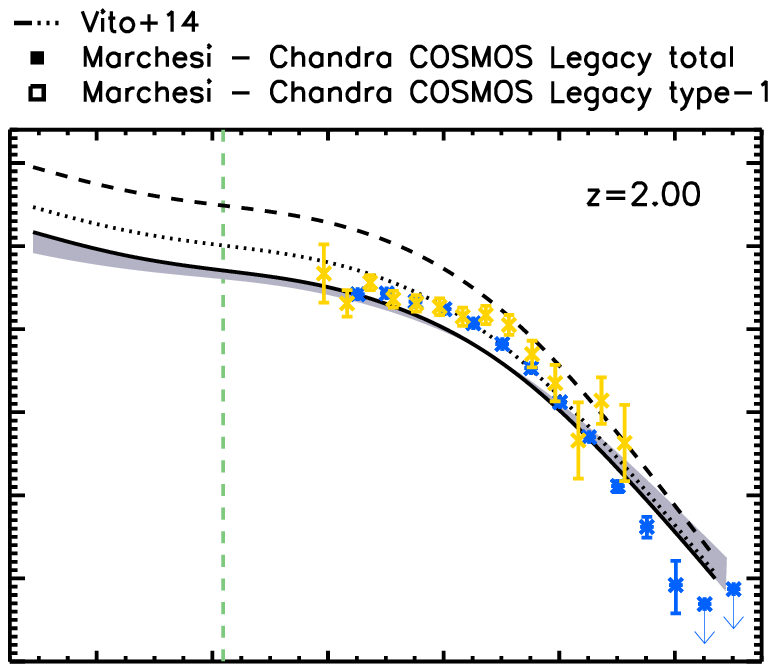}}\vspace{-1.95cm}
	
	\includegraphics[width=1.1\columnwidth]{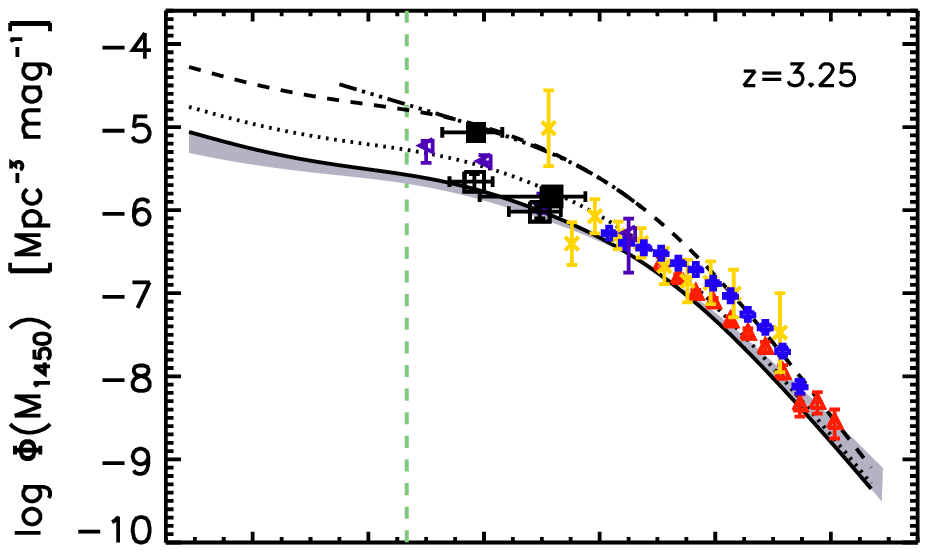}
	\hspace{-2.75cm}{\includegraphics[width=1.1\columnwidth]{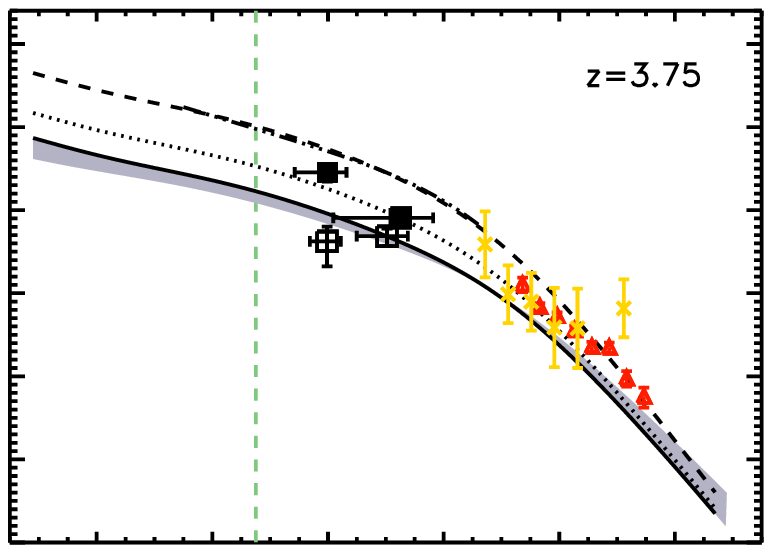}}\vspace{-1.95cm}
	
	\includegraphics[width=1.1\columnwidth]{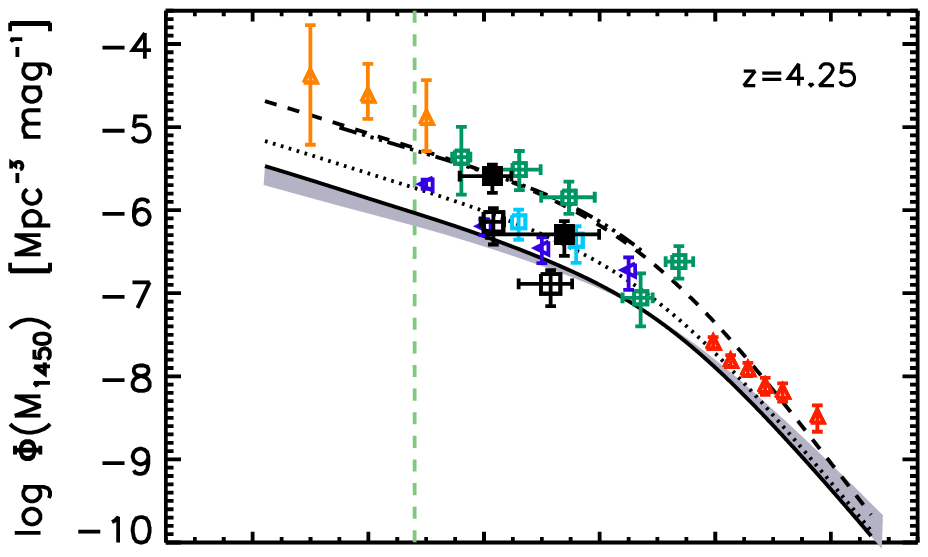}
	\hspace{-2.75cm}{\includegraphics[width=1.1\columnwidth]{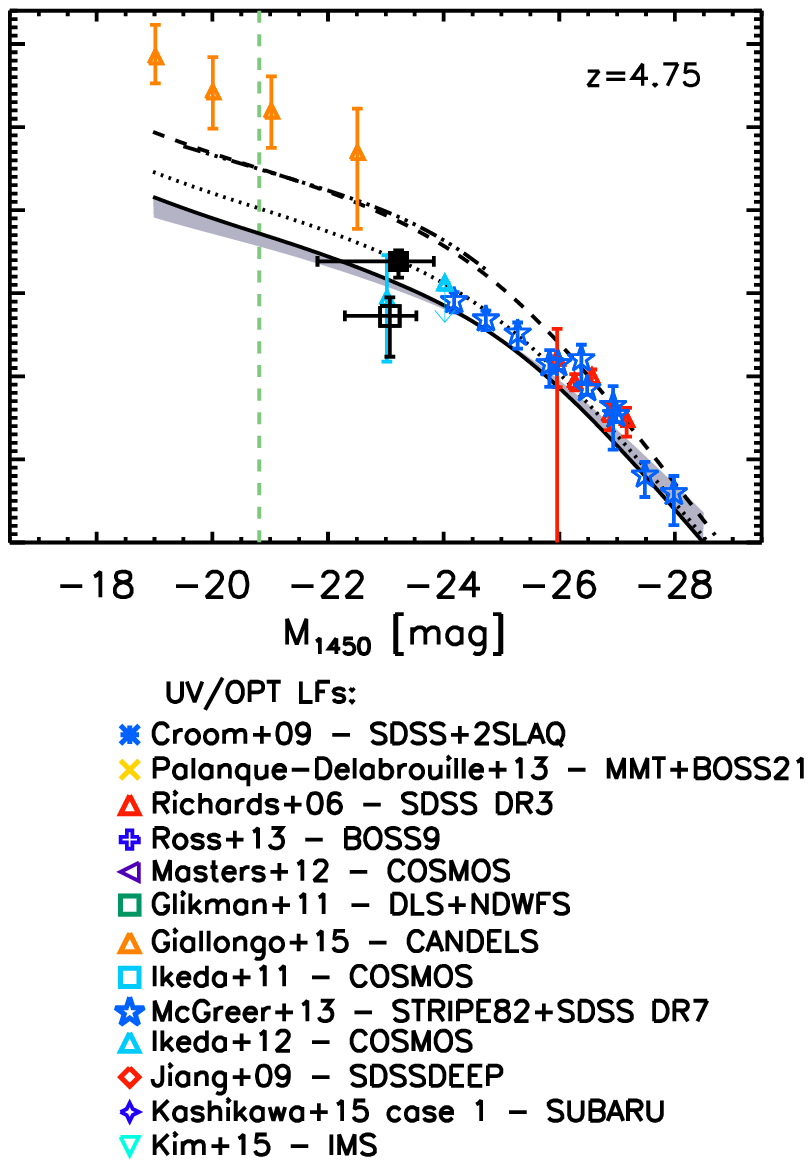}}\vspace{-1.95cm} 
	
	\hspace{-6.63cm}{\includegraphics[width=1.1\columnwidth]{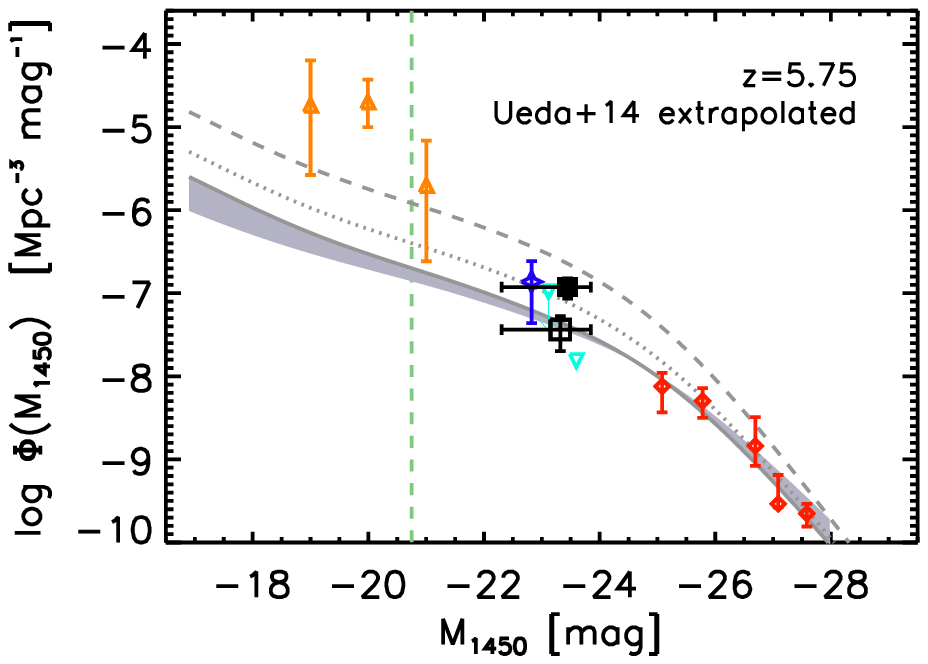}}
	
	\caption{ 
		AGN LFs at different redshifts as a function of the absolute AB magnitude $M_{1450}$. 
		Symbols and colors used for the optical/UV samples are  
		reported in the legend (for more details, see Sect. \ref{sec:QLF}).	
		At $z>3$ we show the new measure of the 2--10 keV LF of type-1 (black open squares) 
		and the whole AGN sample (black filled squares) made by the {\it Chandra} COSMOS Legacy Survey 
		\citep{Marchesi2016b}, and converted into UV magnitudes (see Sect. \ref{sec:uvx}). 
		The plot also shows other recent estimates of the 2--10 keV LFs (converted into UV magnitudes) 
		of unobscured AGN, i.e. ${ \log \rm N_H} < 21$ cm$^{-2}$ \citep[][black solid line]{Ueda2014} 
		and ${\log \rm N_H} < 22$ $cm^{-2}$ \citep[][black dotted line]{Ueda2014}, and inclusive of Compton Thick AGN 
		\citep[][black dashed and triple-dot-dashed lines, respectively]{Ueda2014,Vito2014}. 
		The grey shaded area indicates the effect of changing the $L{\rm _X}-L_{2500}$ (see text for details) 
		relation on the X--ray $\log$N$_{\rm H}<21$ cm$^{-2}$ LF. The X--ray luminosity functions 
		are drawn in grey when they have been extrapolated above existing data. 
		The green dashed vertical line represents the evolution of the break magnitude 
		$M_*$ at 1500 \AA\ of the galaxy UVLF \citep{Parsa2015}.}
	\label{fig:uvlf}
\end{figure*}

All the optical/UV QSO LFs were converted into AB absolute magnitude at 1450 \AA, $M_{1450}$, using 
the expressions
$M_i(z=2)=M_g(z=2)-0.25$ and $M_i(z=2)=M_{1450}(z=0)-1.486$ 
\citep[see, ][eq. 8-9]{Ross2013}, and are shown in Fig. \ref{fig:uvlf} in seven representative redshift bins. 

\begin{itemize}
	
	\item [-] SDSSQS. 
	In the redshift range $0.3<z<5.0$ we use the absolute $i$-band (7470 \AA) binned QLF from the
	Sloan Digital Sky Survey Data Release 3 \citep[SDSS DR3,][]{Richards2006}.
	The sample consists of 15343 QSOs and extends from $i=15$ to 19.1 at $z\la$3 and to $i = 20.2$ at $z\ga$3.	
	
	\item [-] SDSS-2SLAQ. 
	The 2dF-SDSS LRG And QSO survey 
	\citep[2SLAQ,][]{Croom2009} 
	at $0.4<z<2.6$ has 12702 QSOs with 
	an absolute continuum limiting magnitude of $M_g(z = 2) < -21.5$.  	
	
	\item [-] SDSS-III/BOSS QSO survey (DR3).  
	For $0.68<z<4$ down to the limiting extinction corrected magnitude 
	$g = 22.5$, \citet{Palanque-Delabrouille2013} used 
	variability-based selection 
	to measure the QLF\footnote{We adapted their QLF to our adopted cosmology.}.
	The targets were shared between 
	SDSS-III: BOSS (BOSS21) and the MMT, yielding 
	a total of 1877 QSOs. 
		
	\item [-] SDSS-III/BOSS QSO survey (DR9).
	The optical QLF in the range $2.2 <z < 3.5$ 
	has been studied also by \citet{Ross2013}, who 
	targeted $g < 22$ QSOs
	in the BOSS DR9 footprint, achieving 
	a total of 23301 QSOs 
	sampled in the absolute magnitude $-30\leq M_i \leq-24.5$.
	
	\item [-] COSMOS-MASTERS+12.
	The rest-frame UV QLF in the 
	Cosmic Evolution Survey (COSMOS) 
	at $3.1<z<3.5$ and $3.5<z<5$
	was investigated by \citet{Masters2012},
	that reached the limiting apparent magnitude of 
	$I_{AB} = 25$. This sample of 155 likely type-1 AGN is 
	highly complete 
	above $z = 3.1$ in the HST-ACS region 
	of COSMOS.
	
	\item [-] COSMOS-IKEDA+11.
	The same area of the COSMOS field was studied also by \citet{Ikeda2011} 
	in order to probe the faint-end 
	of the QLF at $3.7 \lesssim z \lesssim 4.7$. They reached
	5$\sigma$ limiting
	AB magnitudes $u^* = 26.5$, $g' = 26.5$, $r' = 26.6$, and $i' = 26.1$. 
	They selected 31 QSO candidates using colors ($r'- i'$ vs $g'-r'$) and found 8 spectroscopically confirmed QSOs at $z\sim 4$.	

	\item [-] COSMOS-IKEDA+12.
	\citet{Ikeda2012} searched in a similar way candidates of low-luminosity QSO at $z\sim5$ 
	using the colors $i' - z'$ vs $r'- i'$. Their spectroscopic campaign confirmed 1 type-2 AGN at 
	$z\sim5.07$ and set upper limits on the QLF.	

	\item [-] DLS-NDWFS.
	\citet{Glikman2010} developed a color-selection ($R-I$ vs $B-R$) using 
	simulated QSO spectra. This technique was then used by 
	\citet{Glikman2011} to build the $z\sim4$ UV LF using parts of the Deep Lens Survey (DLS) 
	and NOAO Deep Wide-Field Survey (NDWFS), finding 24 QSOs with $3.74<z<5.06$, down to $M_{1450}=-21$ mag.

	\item [-] SDSS STRIPE82-MCGREER+13. 
	At $4.7<z<5.1$ we use the QLF
	investigated by \citet{McGreer2013}, whose  
	sample has a total of 52 AGN 
	at a limiting magnitude of $i_{AB}=22$.	

	\item [-] SDSS STRIPE82-JIANG+09. 
	\citet{Jiang2009} discovered
	six QSOs at $z\sim6$, four of which
	comprise
	a complete flux-limited sample at $21 <z_{AB} < 21.8$.
 	
	\item [-] SUBARU HIGH-$Z$ QSO survey.
	The SUBARU high-$z$ QSO survey provided an estimate of
	the faint end of the QLF at $z\sim6$.
	\citet{Kashikawa2015}  
	have a sample of 17 QSO candidates at limiting magnitude $z_R < 24.0$,  
	but for 10 of them do not have spectroscopic follow-up, therefore 
	their faintest bin might be a lower limit on the QLF. 

	\item [-] CANDELS GOODS-S.
	The CANDELS GOODS-S field has yielded 
	22 AGN candidates at $4<z<6.5$, five of which 
	have spectroscopic redshifts, 
	down to a mean depth of $H=27.5$ \citep{Giallongo2015}.
	The resulting UV LF lies in the absolute magnitude interval $-22.5 \la M_{1450} \la -18.5$. 	

	\item [-] IMS-SA22.
	Recently \citet{Kim2015} searched for high-$z$ QSOs in one field (i.e. SA22) of the 
	Infrared Medium-deep Survey (IMS). 
	The reached J-band depth corresponds at $z = 6$ to $M_{1450} \simeq -23$ mag. 
	They found a new spectroscopically confirmed QSO at $z = 5.944$, 
	and other six candidates using color selection.\\

\end{itemize}

The Fig. \ref{fig:uvlf} also reports as a green vertical dashed line the evolution of the break magnitude $M_*$ at 1500 \AA\ of the galaxy UVLF, where 
\begin{equation}
M_* = (1+z)^{0.206}(-17.793+z^{0.762})~,
\end{equation}
\citep{Parsa2015}.
The $L_*$ of the galaxy LF indicates the luminosity range where galaxy contribution to the ionizing background could be relevant.
Indeed the galaxy number density at $M_*$ is much higher than the QSO 
one of at least two orders of magnitudes at any redshift. 
	For example, at redshift 0.90 (4.25) the galaxy density at the break luminosity 
	$M_* = -19.2$ $(-20.8)$ result to be $\sim 8\times10^{-4}$ ($\sim 4\times10^{-4}$) Mpc$^{-3}$ mag$^{-1}$ 
	\citep{Parsa2015}. 
Therefore galaxies could play a leading role in the \HI\ reionization at early epochs, 
even with little $f_{esc}$.

\subsection{X-ray Luminosity Functions}\label{sec:XLF}

We have based our analysis on the \citet{Ueda2014} XLF, which is obtained using 4039 sources 
from 13 different X--ray surveys performed with \textit{Swift}/BAT, \textit{MAXI}, \textit{ASCA}, 
\textit{XMM-Newton}, \textit{Chandra} and \textit{ROSAT}. These sources have been detected in 
the soft (0.5--2 keV) and/or hard ($>2$ keV) X--ray bands. The 2--10 keV LF has been computed 
in the redshift range 0$<z<$5 and in the luminosity range 42$<\log L\rm_X<$46.5 erg s$^{-1}$. 
The faint end of this range is luminous enough (i.e., an order of magnitude larger) 
to exclude the contribution to the X--ray emission of both X--ray binaries and hot 
extended gas \citep[see, e.g.][]{Lehmer2012,Basu-Zych2013,KimFabbiano2013,Civano2014}.
Indeed, the threshold of $\sim 10^{41}$ erg s$^{-1}$ translates, according to 
\citet[][see their eq. 12]{Lehmer2012}, into a SFR of $\sim$ 10-100 M$_{\odot}$ yr$^{-1}$.

\citet{Ueda2014} found that the shape of the XLF significantly changes with redshift: in the local Universe, the 
faint-end slope (i.e., below the XLF break) is steeper than in the redshift range 1$<z<$3.
\citet{Ueda2014} computed the XLF in various absorption ranges (i.e., at different $\rm N_H$): they confirmed the existence 
of a strong anti-correlation between the fraction of absorbed objects and the 2--10 keV luminosity. At high luminosities, 
the majority of AGN  are unabsorbed, while moving to low luminosities ($\log L\rm_X<$43.5 erg s$^{-1}$ 
in the redshift range 0.1$<z<$1) the contribution of absorbed AGN to the XLF becomes dominant. 
Moreover this trend evolves with redshift, maintaining the same slope but shifting toward 
higher luminosities \citep[see also][]{LaFranca2005,Hasinger2008}.

\citet{Vito2014} computed the 2--10 keV AGN LF in the redshift 
range $3< z \leq5$ from a sample of 141 sources selected in 
the 0.5--2 keV band. The sample was obtained combining four different 
surveys down to a flux limit $\sim$ 9.1 $\times$ 10$^{-18}$ erg s$^{-1}$ 
cm$^{-2}$. In this redshift range the XLF is well described by a pure 
density evolution model, i.e., there is no luminosity dependence on the 
shape of the LF at different redshifts.
In this work, the whole 
	XLF by \citet{Vito2014} has been used without separating according to 
	the N$_{ \rm H}$ classification.

The \textit{Chandra} COSMOS-Legacy \citep{Civano2016,Marchesi2016} 
$z\geq3$ sample \citep{Marchesi2016b} contains 174 sources with 
$z\geq3$, 27 with $z\geq$4, nine with $z\geq$5, and four with $z\geq$6. 
87 of these sources have a reliable spectroscopic redshift, 
while for the other 87 a photometric redshift has been computed \citep{Salvato2011}.
The photo-$z$ mean error is $\sim$5-10\%, 
with 60\% of the sample having uncertainties less than 2\% 
while for 10\% of the sample the uncertainties are greater than 20\%.
The 2--10 keV \textit{Chandra} COSMOS-Legacy $z\geq3$ sample 
is complete at $L\rm_{X}>10^{44.1}$ erg s$^{-1}$ 
in the redshift range 3$<z<$6.8; 
at lower luminosities (10$^{43.55}<L_{\rm X}<$10$^{44.1}$ erg s$^{-1}$) 
the sample is complete in the redshift range 3$<z<$3.5. 
For the purposes of this work, we computed the space densities also at $z$=3.75, 
in the luminosity range 10$^{43.7}<L_{\rm X}<$10$^{44.1}$ erg s$^{-1}$, 
and at $z$=4.25, in the luminosity range 10$^{43.8}<L_{\rm X}<$10$^{44.1}$ erg s$^{-1}$.
The \textit{Chandra} COSMOS-Legacy $z\ge3$ sample has also been divided in two subsamples: 
one is made by 85 type-1 AGN, 
on the basis of their spectroscopical classification, 
or, when only photo-$z$ was available, with a Spectral Energy Distribution (SED) fitted with an unobscured AGN template; the second subsample is formed by the 89 optically-classified type-2 AGN, either without evidence of broad lines in their spectra, or SED best-fitted by an obscured AGN or a galaxy template 
\citep[for the description of the optical sources classification see][]{Marchesi2016}.

\section{Comparison between UV/optical and X--ray LFs}\label{sec:uvx}

To compare the XLFs 
to the UV/optically-selected QLFs, we converted the X--ray luminosities into UV ones.
A X--ray photon index $\Gamma = 1.8$ was used to compute the monochromatic 2 keV luminosity $L\rm_{2keV}$,
which was then converted into 2500 \AA\ luminosity $L\rm_{2500}$ 
using
\begin{equation}
 \log L_{2500} = (1.050 \pm 0.036) \log L_{\rm 2keV} + (2.246 \pm 1.003) \, ,
\end{equation}
obtained inverting eq. 5 of \citet{Lusso2010} where the UV luminosity was treated as the dependent variable.
Both monochromatic luminosities are in erg s$^{-1}$ Hz$^{-1}$.
\citet{Georgakakis2015} used a similar approach to derive the 
UV LF, but they adopted eq. 6 of \citet{Lusso2010}, which is the bisector best fitting.
However, as we are interested in predicting the UV luminosity starting from the X--ray data,
the relation where the $L_{\rm 2500}$ is a function of $L_{\rm 2 keV}$ was preferred.
In order to correctly reproduce the UV/optically-selected LF, 
a redshift-independent observed spread of $\sim0.4$ dex was applied \citep{Lusso2010}.
This value takes into account intrinsic dispersion, variability, and measurement uncertainties.
 
A power-law SED $L_{\nu}\propto \nu^{-\alpha_{\nu}}$ \citep[e.g. following][]{Giallongo2015} with $\alpha_{\nu}=0.44$ for $1200<\lambda<5000$ \AA\ \citep{Natali1998,VandenBerk2001} and
$\alpha_{\nu}=1.57$ when $228<\lambda<1200$ \AA\ \citep{Telfer2002} was adopted to obtain the UV luminosity $L\rm_{1450}$.
Finally we converted $L\rm_{1450}$ into AB absolute mag $M\rm_{1450}$ using
\begin{equation}
\label{eq:LM}
{L_{\rm1450} = 4 \pi d^2 10^{-0.4M_{\rm1450}}} f_0~,
\end{equation}
where $d=10~ {\rm pc}=3.0857\times10^{19}$ cm and $f_0=3.65\times10^{-20}$ erg cm$^{-2}$ s$^{-1}$ Hz$^{-1}$ is the zero-point.

In Fig. \ref{fig:uvlf} we show the AGN UV LF, as measured by the
optically-selected QLFs, already described in Sect. \ref{sec:QLF}, together with the XLF by 
\citet[][black open squares for type-1 AGN and black filled squares for the whole sample]{Marchesi2016b}.
We also show  
the \citet{Vito2014} XLF (black triple-dot-dashed line) and 
the \citet{Ueda2014} XLF in three different $\rm N_H$ regimes: $\log \rm N_H<21$ cm$^{-2}$ 
(black solid line); $\log \rm N_H<22$ cm$^{-2}$ (black dotted line);
$\log \rm N_H<26$ cm$^{-2}$ (whole AGN population, i.e., including Compton Thick sources, black dashed line). 
All the XLFs have been plotted only in the redshift and X--ray luminosity ranges where 
data exist. 
At $z=5.75$, the \citet{Ueda2014} XLF has been extrapolated (it has been originally computed in the redshift range 0$<z<$5, 
see Sect. \ref{sec:XLF}), and then it has been drawn in grey in Fig. \ref{fig:uvlf}.
The grey shaded region shows the effect of changing the relations between $L_{\rm 2 keV}$ and $L_{\rm 2500}$ in the convertion of the \citet{Ueda2014} 
$\log$N$_{\rm H}<21$ cm$^{-2}$ XLF: beside the 
relation by \citet{Lusso2010},  
the relation found by \citet[][see their eq. 1b]{Steffen2006} was used to obtain the upper and lower limits on this area. 
 
It is worth noticing the perfect agreement between 
\citet{Vito2014} and \citet{Ueda2014} XLF having $\log \rm N_H < 26$ cm$^{-2}$ 
over the whole luminosity range in which both XLFs exist and in all the redshift bins 
with
$z>3$, taken into account in our analysis. 
As noted in Sect. \ref{sec:XLF}, in this work the XLF computed by \citet{Vito2014} has not been 
	divided in $\rm{N_H}$ classes and then describes the whole X--ray emitting AGN population. 
Also the new measures from the {\it Chandra} COSMOS Legacy Survey $z>3$ sample are in good agreement with the above two XLFs at all redshifts, given the current uncertainties both in X--ray luminosity 
and density.
The most important contribution coming from the \textit{Chandra} COSMOS Legacy Survey $z>3$ sample 
is that it confirms the extrapolation of the \citet{Ueda2014} XLF at luminosities 
$L_{\rm X} \sim L_*$ and $z>5$ 
(see Fig. \ref{fig:uvlf}),
therefore 
supporting the 
use of the \citet{Ueda2014} XLF also for $z\sim5-6$, 
which is one of the key epochs 
for reionization studies.
Given the good agreement between these different X--ray LFs, 
independently computed, 
we claim that the AGN description coming from the X--ray selected samples is coherent and robust, 
clearly confirming the global ``downsizing'' evolution, 
where more luminous AGN have their number density peak at higher redshifts compared with less luminous ones.

As shown in Fig. \ref{fig:uvlf}, there is a fairly good agreement between the UV/optical binned QLFs and 
the 2--10 keV $\log$N$_{\rm H}$<21 and $\log$N$_{\rm H}$<22 cm$^{-2}$ AGN LFs up to $z\sim6$, in the luminosity range of the break and beyond 
(i.e., $M_{1450} \leq -23$).
As expected, this result is in agreement with the unification model where (as discussed in the Introduction) 
the X--ray $\log$N$_{\rm H}\lesssim21-22$ cm$^{-2}$ AGN population should correspond to the UV/optically-selected QSOs \citep[see also][]{Cowie2009}.
We note that this matching between optical QSOs and X--ray $\log$N$\rm{_H}\lesssim21-22$ cm$^{-2}$ 
population up to $z=6$ is also in agreement with what was found by \citet{Risaliti2015}, who showed that $\alpha_{OX} (L_X)$ is redshift-independent, and can therefore be used 
to derive a
cosmological distance indicator. 
Indeed this matching implies, as we have assumed, that a change of the relation between
$L_{\rm2500}$ and  
$L_{\rm2keV}$ with redshift is not required.

\begin{figure*}
	\centering
	
	\includegraphics[width=1.1\columnwidth]{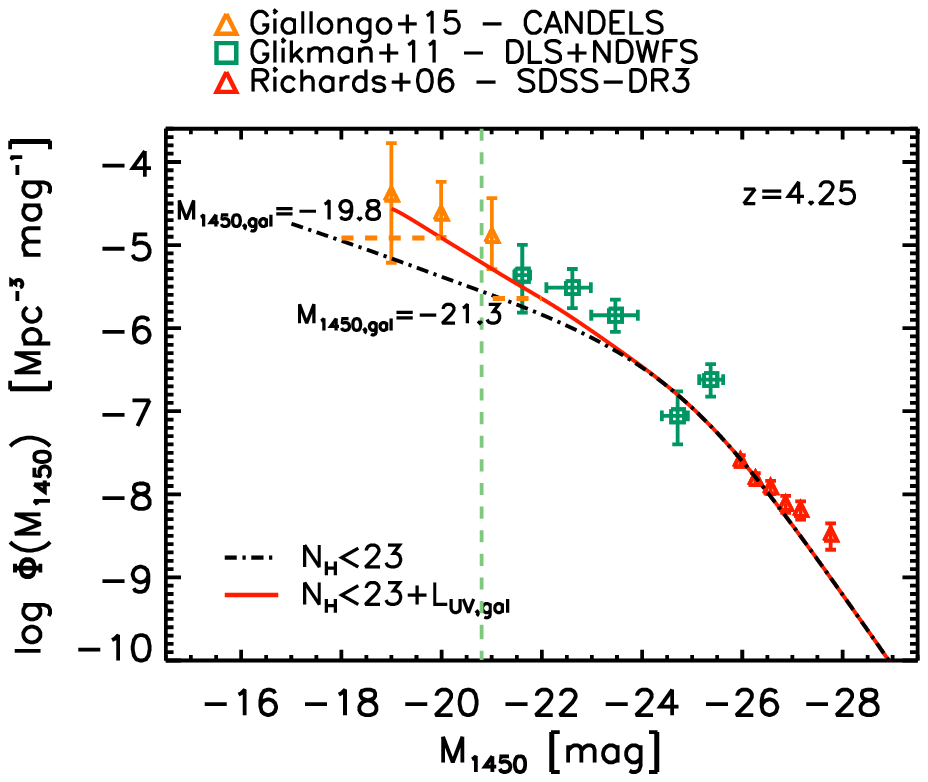}
	\hspace{-2.75cm}{\includegraphics[width=1.1\columnwidth]{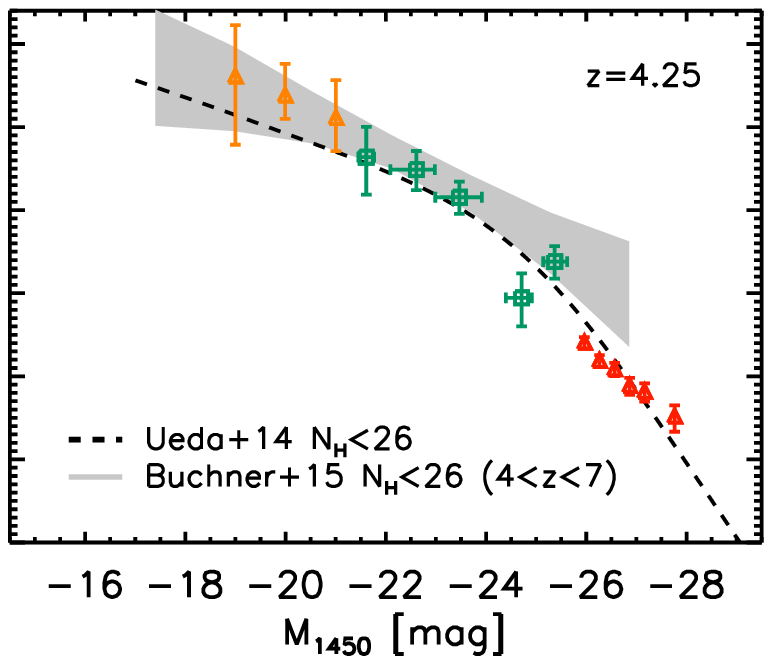}}
	
	\caption{UV LF at redshift 4.25. 
		\textit{Left:} The faint end of the UV LF measured by UV/optically selected samples 
		of \citet[][orange triangles]{Giallongo2015} and \citet[][green squares]{Glikman2011} 
		could be described by the $\log \rm N_H<$23 cm$^{-2}$ XLF of \citet{Ueda2014} 
		adding a UV contribution arising from 
		the luminosity of the host galaxy with magnitudes in the range $-21.5<M_{1450}<-19.5$,
			which are typical of the galaxy LF break luminosity at this redshift \citep[$M_{1500}\sim -20.8$, green dashed vartical line, ][]{Parsa2015}.	
			A Gaussian convolution that accounts for the observed spread ($\sigma\sim0.4$) in the 
			relation $L_X - L_{2500}$ (red line) has also been included.
		\textit{Right:} Comparison of the UV/optically selected samples and two different 
		XLFs that consider also the contribution of the Compton Thick AGN ($\log \rm N_H<$26 cm$^{-2}$), namely the \citet[][black dashed line]{Ueda2014} and \citet[][grey shaded area]{Buchner2015}.}
	\label{fig:faintUV}
\end{figure*}

\subsection{Faint-end of the UV LF at $z>4$}\label{sec:faintUV}
The determination of the AGN LF faint end is one of the still open 
problems in extragalactic astronomy, and it translates into a poor knowledge of the AGN 
demography and evolution, specially when moving at $z>4$.

At luminosities lower than the break (i.e., $M_{1450} \geq -23$), a good agreement 
between the UV/optical and X--ray samples is found only up to redshift 
$\sim4$, while at higher redshifts, 
the $\log$N$_{\rm H}\lesssim$22 cm$^{-2}$ XLF of \citet{Ueda2014} underpredicts up to a factor of $\sim$ 1 dex the 
LFs by \citet{Glikman2011} and \citet{Giallongo2015}, which were both measured using UV-restframe selected samples.
This disagreement between X--ray and rest-frame 
	UV selected AGN samples has been recently found also by \citet{Vito2016}, 
	who derived an upper limit on the AGN XLF by stacking the X--ray counts 
	in the CDF-S 7Ms, which is the deepest X--ray survey to date.
In what follows we discuss a few possible scenarios that may explain the origin of these discrepancies.

As already discussed by \citet{Giallongo2015}, 
	it could be possible that in the UV fluxes of their sample 
	the contribution of the stellar emission of the hosting galaxy is not negligible.
	Indeed, as shown in Fig. \ref{fig:faintUV} (left panel, red solid line) 
	the $\log \rm N_H < 23$ cm$^{-2}$ XLF\footnote{This XLF has 
			been convolved with a 0.4 dex Gaussian scatter as already described in Sect. \ref{sec:uvx}.} 
	(corresponding to the typical $\rm N_H$ value 
	measured in the sample of \citealt{Giallongo2015}), 
	could reproduce 
	the \citet{Glikman2011} and \citet{Giallongo2015} measures 
	once typical $L_*$ galaxy luminosities 
	($-21.5<M_{1450}<-19.5$ at $z=4.25$, orange dashed horizontal lines in Fig. \ref{fig:faintUV})
	are added to the AGN luminosity. 
	The left panel of Fig. \ref{fig:faintUV} also shows for guidance 
	the break magnitude at 1500 \AA\ of the galaxy UVLF \citep[green dashed vertical line, ][]{Parsa2015}.
A non-negligible galaxy contribution 
can be also related to 
a scenario in which 
the faint (not heavily absorbed) AGN population with $\log$N$_{\rm H} \lesssim 23$ cm$^{-2}$
could, through outflows and mechanical feedback,  
increase the 
$f_{esc}$ of SFGs by cleaning the environment and enhancing the porosity 
of the ISM \citep[see, e.g.][]{Giallongo2015,Smith2016}.
This mechanism could be rather effective at higher redshift as 
the UV emission could be produced more efficiently by Pop III stars 
\citep{Kimm2016}.

An alternative scenario is that at high redshift the UV-selected samples are not only
associated to 
X--ray $\log \rm N_H\lesssim$22 cm$^{-2}$ AGN but also to heavily X--ray absorbed AGN ($\log$N$_{\rm H}$>25 cm$^{-2}$).
If the AGN population which 
mostly contribute to the ionizing background have greater $\rm N_H$ 
values than that considered in our minimal baseline model (which have been chosen to
match the optical/UV LFs at $z<4$), then the $f_{esc}$ does not become zero rapidly for $\rm N_H>10^{22}$ cm$^{-2}$.
Therefore, the AGN contribution to UV emission would be enhanced. 
Indeed, in this case the underprediction by the $\log$N$_{\rm H}<$26 cm$^{-2}$ XLF  
of the UV LFs reduces to $\lesssim$ 0.5 dex.
This scenario is rather unlikely. 
	Indeed, although there is evidence that the 
	ratio between hydrogen column density and 
	extinction in the V band
	(i.e. the 
	$\rm{N_H / A_V}$ ratio)
	could be larger than Galactic in AGN \citep[see e.g.][]{Maiolino2001, Burtscher2016},
	it has been shown that 
	all AGN having $\rm{N_H>10^{22}}$ cm$^{-2}$ show 
	$A_V>5$ \citep{Burtscher2016}, 
	therefore the UV photons will be likely absorbed 
	within the host galaxy and do not contribute to
	the cosmic ionization.
	Moreover \citet{Burtscher2016} showed  that if accurate X--ray and 
	optical analysis is carried out, an agreement between the X--ray and 
	optical classification is found.

Recently it has been claimed that the
XLF \citep{Buchner2015,Fotopoulou2016} 
has a higher density of low-luminosity AGN 
compared to the estimates from \citet{Ueda2014, Miyaji2015, Aird2015,Aird2015Nustar, Georgakakis2015, Vito2014}.
The right panel of Fig. \ref{fig:faintUV} shows the whole AGN XLF (i.e. including the Compton Thick population) by \citet{Buchner2015}, 
converted into UV magnitudes (grey shaded area),
together with the corresponding XLF by \citet[][dashed black curve]{Ueda2014}
and the UV/optically selected samples at $z=4.25$.
The two XLFs were converted into UV according to our standard procedure highlighted in Sect. \ref{sec:uvx}.
Concerning the X--ray population with $\log \rm N_H<26$ cm$^{-2}$, the XLF by \citet{Buchner2015}
is indeed $\sim$ 1 dex higher than the XLF by \citet{Ueda2014} and this 
factor is enough to reproduce the UV/optically selected samples. 
Nonetheless, it must be considered that this agreement 
is found only if the heavily X--ray absorbed AGN population is also included.  
The uncertainties are quite large ($\sim$1 dex in density and the redshift bin is very broad, $4<z<7$), 
and
\citet{Buchner2015} state that 
this is the reason for possibly not finding
a steep decline with redshift in their AGN space density. 

In summary, we found that there is a discrepancy at the  
faint-end of the UVLF between the measures derived using direct UV data
and the prediction of the X--ray $\log \rm N_H\lesssim$22 cm$^{-2}$ AGN LF at $z>4$.
This discrepancy, which is absent at lower redshifts, can be attributed either to a small 
contribution of UV emission in the AGN host galaxy 
of the order of the galaxy UV $L_*$
or, unlikely, to a
substantial contribution to the UV background 
coming from an X--ray absorbed population 
whose density and escape fraction, consequently, should have
been underestimated by most of the previous 
studies.

However, inside the AGN unified models 
it should be remembered (see Introduction) that, for isotropic 
reasons, we are interested in the 
unabsorbed
$\log$N$_{\rm H}<22$ cm$^{-2}$ population
as measured along the line-of-sight
since the fraction of ionizing AGN should be the same for any observer in the Universe.

\section{QSO UV emissivity}\label{sec:emi}
As discussed at the end of Sect. \ref{sec:uvx}, the good agreement between the optical QSO LF 
and the $\log$N$_{\rm H}\lesssim21 - 22$ cm$^{-2}$ XLF 
indicates that this XLF is a good proxy to estimate the space density of ionizing AGN (i.e., the QSOs).
Therefore we will assume that the $\log$N$_{\rm H}<21$ cm$^{-2}$ AGN have
$\langle f_{esc} \rangle=1$, while the rest of the population has rapidly decreasing $f_{esc}$ as $\rm N_H$ increases.

Most of the previous studies on the estimate of the ionizing background
produced by QSOs have used UV/optically-selected sample and, 
as shown in Fig. \ref{fig:uvlf}, they needed
to extrapolate the QLF below the break luminosity 
\citep[see, e.g. ][]{KS2015,MH2015}. 
On the contrary, the use of the $\log$N$\rm{_H}<$21 cm$^{-2}$ XLF allows us to 
measure the QSO emissivity without any extrapolation at low luminosities
down to five magnitudes fainter than optical surveys 
and up to $z\sim 5$.

In order to investigate the contribution of QSO to the \HI\ ionizing background, 
we calculated the 1 ryd comoving emissivity
\begin{equation}
\label{eq:emi}
\epsilon_{912}(z)=\langle f_{esc} \rangle \int_{ L_{\rm min}} \Phi(L_{912},z)\, L_{912}\, dL_{912} ~,
\end{equation}
where 
$\langle f_{esc} \rangle$ is the mean value of escaping fraction of UV photons, 
$L_{\rm 912}$ is the monochromatic luminosity at 912 \AA, $\Phi(L_{\rm912},z)$ is the QLF 
and $L_{\rm min}$ sets the lower limit for the luminosity integration.
We converted the $M_{\rm1450}$ into $L_{\rm912}$ using our adopted SED, as described in Sect. \ref{sec:uvx}.

\begin{figure*}
	\centering
	\includegraphics[width=1.1\columnwidth]{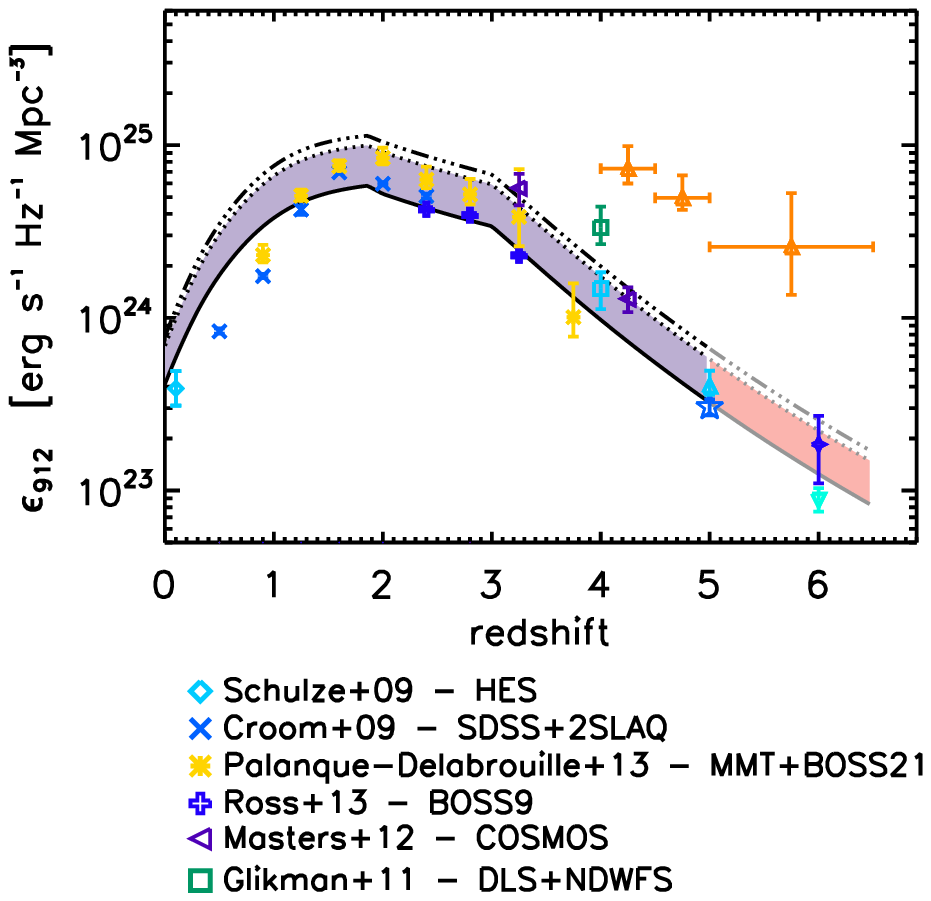}
	\hspace{-2.75cm}{\includegraphics[width=1.1\columnwidth]{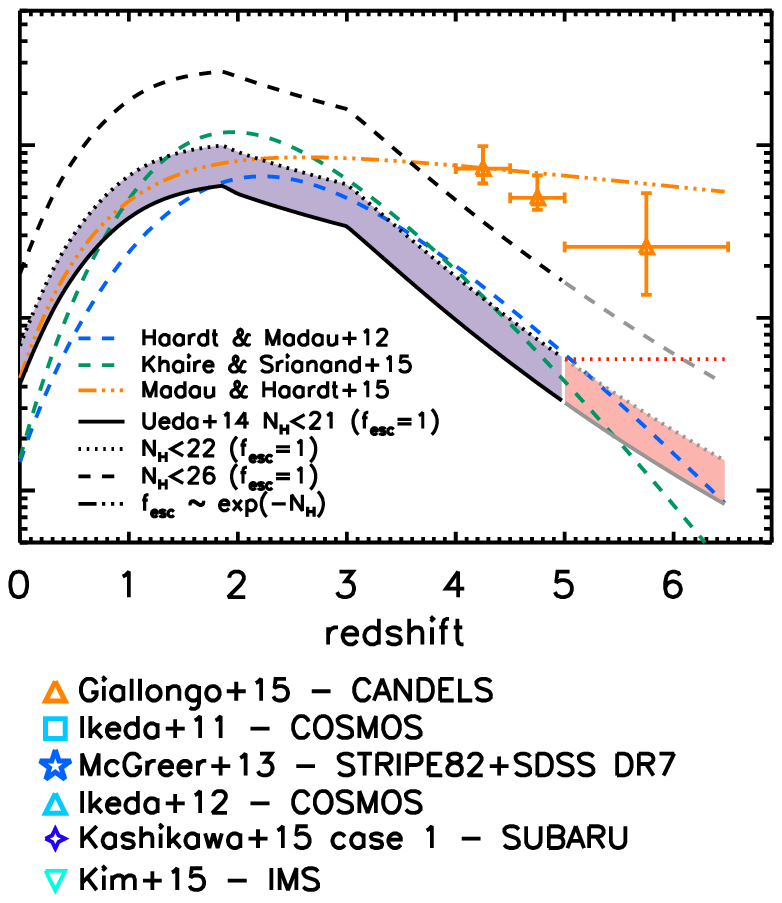}}
	
	\vspace{0.6in}
	\caption{Redshift evolution of the hydrogen ionizing emissivities, $\epsilon_{912}(z)$. \textit{Left:} 
		$\epsilon_{912}$ computed using the UV/optical binned QLF described in Sect. \ref{sec:QLF}, colors and symbols are reported in the legend.  
		The solid and dotted black lines are the $\epsilon_{912}$ computed 
		(with $f_{esc}=1$) from the 2--10 keV LF by \citet[][solid line for the $\log \rm N_H<21$ cm$^{-2}$, black dotted for $\log \rm N_H<22$ cm$^{-2}$]{Ueda2014}. 
		The triple-dot-dashed black line shows the resulting 
			emissivity computed assuming $f_{esc}\sim e^{-\rm N_H}$ (see text for more details).
		The emissivities are drawn in grey when XLFs are extrapolated. The shaded area shows our best estimate of the 
		UV ionizing AGN emissivity, which should lie in between the two limits of $\log$N$_{\rm H}<21$ and $\log$N$_{\rm H}<22$ cm$^{-2}$. When the XLFs are extrapolated the shaded area is plotted in pink.	
		\textit{Right:} 
		The black dashed line is $\epsilon_{912}$ computed from the 2--10 keV LF by \citet{Ueda2014} inclusive of Compton Thick sources. The red dotted horizontal line shows the 
		upper limit on the $\log \rm N_H<22$ population assuming that 
		the XLF remains constant for $z>5$.
		The other curves are the prediction of the evolution of the emissivity with redshift from \citet[][blue dashed]{HM2012}, \citet[][green dashed]{KS2015} and \citet[][orange triple-dot-dashed]{MH2015}.}
	\label{fig:emissivity}
\end{figure*}

\subsection{Comparing optical and X--ray emissivities}\label{sec:emiuvx}
Figure \ref{fig:emissivity} (left panel) shows the evolution of the comoving ionizing emissivity $\epsilon_{912}$ 
as a function of redshift for the $\log$N$_{\rm H}$<21 and $\log$N$_{\rm H}$<22 cm$^{-2}$ X--ray population 
(solid and dotted black lines, respectively). 
When the XLF has been extrapolated, i.e. at $z>5$, the emissivities are drawn in grey.
The above computed X--ray emissivities are reported in Table \ref{tab:emi} up to $z=7$.

\begin{table}
	\centering
	\caption{Redshift evolution of emissivities at 912 \AA\ obtainted integrating 
		the XLFs of \citet{Ueda2014} with Eq. \ref{eq:emi} with $\log L_{\rm min}=$27.22 erg s$^{-1}$ Hz$^{-1}$. 
		Columns are: (1) redshifts, (2) logarithm of the 
		\HI\ ionizing emissivity computed from the X--ray $\log \rm N_H<$21 cm$^{-2}$ AGN,
		(3) same as column (2) but for $\log \rm N_H<$22 cm$^{-2}$ AGN.
		Both columns (2)-(3) have been computed with $\langle f_{esc} \rangle =1$ 
		and are in units erg s$^{-1}$ Hz$^{-1}$ Mpc$^{-3}$.}
	\label{tab:emi}
	\begin{tabular}[h]{ccc} 
		\hline
		$z$ & \multicolumn{2}{c}{$\log \epsilon_{912}$}\\
		  & ($\rm N_H <10^{21}$)& ($\rm N_H <10^{22}$) \\
		(1) & (2) & (3) \\  
		\hline
       0.0  &  23.62   &  23.84  \\
       0.2  &  23.91   &  24.14  \\
       0.4  &  24.15   &  24.39  \\
       0.6  &  24.33   &  24.57  \\
       0.8  &  24.47   &  24.71  \\
       1.0  &  24.58   &  24.82  \\
       1.2  &  24.65   &  24.90  \\
       1.4  &  24.70   &  24.94  \\
       1.6  &  24.74   &  24.97  \\
       1.8  &  24.76   &  24.99  \\
       2.0  &  24.72   &  24.96  \\
       2.2  &  24.67   &  24.92  \\
       2.4  &  24.64   &  24.88  \\
       2.6  &  24.60   &  24.84  \\
       2.8  &  24.56   &  24.81  \\
       3.0  &  24.53   &  24.77  \\
       3.2  &  24.42   &  24.67  \\
       3.4  &  24.31   &  24.56  \\
       3.6  &  24.20   &  24.45  \\
       3.8  &  24.09   &  24.34  \\
       4.0  &  23.99   &  24.24  \\
       4.2  &  23.89   &  24.14  \\
       4.4  &  23.79   &  24.04  \\
       4.6  &  23.70   &  23.94  \\
       4.8  &  23.60   &  23.85  \\
       5.0  &  23.51   &  23.76  \\
       5.2  &  23.42   &  23.67  \\
       5.4  &  23.34   &  23.59  \\
       5.6  &  23.25   &  23.51  \\
       5.8  &  23.17   &  23.42  \\
       6.0  &  23.09   &  23.38  \\
       6.2  &  23.02   &  23.27  \\
       6.4  &  22.95   &  23.20  \\
       6.6  &  22.87   &  23.12  \\
       6.8  &  22.80   &  23.05  \\
       7.0  &  22.73   &  22.98  \\
		\hline
	\end{tabular}
\end{table}

As a comparison we also 
report in the left panel of the same figure other measures 
which we have derived using the UV/optically-selected QSO samples (see Sect. \ref{sec:QLF}).
We set $L_{\rm min}$ as the faintest luminosity bin available in each survey.
In particular, when integrating the XLF we set $\log L_{\rm min}=$27.22 erg s$^{-1}$ Hz$^{-1}$ 
(i.e. $\log L_{\rm X}=42$ erg s$^{-1}$). 
The choice of not extrapolating the LF in a luminosity range not yet sampled by current surveys is conservative and
implies that the derived QSO ionizing emissivities are, strictly speaking, lower limits. 
Indeed, the adoption of the XLF as an
unbiased representation of the UV/optical QSO LF
allows us to extend the lower luminosity limits of the optical LFs 
(see Fig. \ref{fig:uvlf}) and 
then reach fainter $L_{\rm  min}$ in the integration of Eq. \ref{eq:emi}.
When the XLF has been extrapolated, i.e. at $z>5$, we have 
assumed the evolution implied by \citet{Ueda2014} and we have solved 
Eq. \ref{eq:emi} setting $L_{\rm  min}$ as previously done.

Given the definition in Eq. \ref{eq:emi}, the emissivity is proportional to the 
area beneath the curve $L \times \Phi(L)$. Due to the double-power law shape of 
the LF, $ L \times \Phi(L)$ presents a maximum located in the luminosity break region $L_*$.
Therefore at each redshift the leading contribution to emissivity comes from 
AGN at $L_{*}$, and this is true as far as the faint-end slope of the LF is not too steep. 
 
In order to show the emissivity at $z=0$ derived from the optically-selected QLF, we used the $B$-band double power-law QLF from \citet[][cyan diamond with error bars, for the LF see their Tab. 4]{Schulze2009}\footnote{
The Vega absolute B magnitude $M_B$ were converted into $B$-band luminosity $L_B$ 
in a similar way as in Eq. \ref{eq:LM} (substituting the magnitudes and luminosities) with $f_0=4.063\times10^{-20}$ 
erg s$^{-1}$ cm$^{-2}$ Hz$^{-1}$. Finally $L_B$ was translated into $L_{912}$ with our adopted SED (see Sect. \ref{sec:uvx}). The integration limit used was $\log L_{min}=29.42$ erg s$^{-1}$ Hz$^{-1}$.
The uncertainties on this local emissivity 
have been evaluated directly from the uncertainties on the binned QLF.
}. 
At $z>4$ we also show the results from \citet[][orange triangles with error bars]{Giallongo2015}.

The contribution of the X--ray $\log \rm N_H<$21 cm$^{-2}$ population should be considered 
as a lower-limit to the AGN ionizing emissivity (black solid line in Fig. \ref{fig:emissivity}), in fact 
also the $21< \log \rm N_H<22$ cm$^{-2}$ AGN could 
contribute significantly.
Inside our minimal model, an upper limit to the QSO emissivity can be derived (black dotted line in Fig. \ref{fig:emissivity}) 
under the hypothesis of $\langle f_{esc} \rangle=1$
up to $\log \rm N_H=22$ cm$^{-2}$ and then sharply zero for the rest of the AGN.
	We tested how strong is this last approximation
	on the upper limit to the AGN contribution to the emissivity
	by assigning 
	a $f_{esc}$ depending on the column density $\rm N_H$.
	 Indeed 
	 a relation 
	 between the escape fraction and the extinction
	 of the type $f_{esc} \propto e^{-A_V}$
	 is expected \citep{Mao2007}.
	  Therefore in a less simplified (and more realistic) model,
	  assuming a constant $\rm N_H / A_V$ ratio, the escape fraction is expected to
	  exponentially depend on the $\rm N_H$.
	   In this scenario, an escape fraction equal to unity at $\log \rm N_H=21$ cm$^{-2}$
	   will drop to $\sim 0.37$ at $\log \rm N_H=22.5$ cm$^{-2}$ and to 
	   $\sim5\times10^{-5}$ already at 
	   $\log \rm N_H=23.5$ cm$^{-2}$.
	    This calculation is shown in Fig. \ref{fig:emissivity} (left panel) 
	    as a triple-dot-dashed black line. 
	    The emissivity computed 
	    assuming the exponential dependence on $\rm N_H$
	    is only slightly enhanced ($\sim17$\%)
	    with respect to the first simplified
	    approximation of $f_{esc}$ sharply zero for $\rm N_H>10^{22}$ cm$^{-2}$. 
	   An almost identical result is obtained also assuming 
	   $f_{esc}=1$ up to $\log \rm N_H=22$ cm$^{-2}$ and 
	   an additional
	    constant 
	    $f_{esc}=0.1$ for the 
	    $22<\log \rm N_H <26$ cm$^{-2}$ AGN population. 
	   Since the scenario in which 
	   the $f_{esc}$ is dependent on $\rm N_H$
	   does not alter 
	   significantly our initial simplified and rather strong
	   approximation that 
	   the $f_{esc}$ sharply drop to zero for all AGN having $\log \rm N_H>22$ cm$^{-2}$, 
	   we will continue to use as upper limit on the AGN emissivity the 
	   one calculated assuming $f_{esc}=1$ up to $\log \rm N_H=22$ cm$^{-2}$ only.

As expected, the estimates obtained using the X--ray $\log \rm N_H<$21 and $\log \rm N_H<$22 cm$^{-2}$ LFs  
are in good agreement with most of the estimates coming from the 
optical/UV QLFs, where available. 
At variance, as already noticed in Sect. \ref{sec:uvx},
the measures at $z>4$ by \citet{Glikman2011} and \citet{Giallongo2015} 
are up to a factor of $\sim$ 8 larger than the other estimates, obtained both from optical and X--ray data.
This difference was also pointed out 
by \citet{Georgakakis2015} 
who integrated the XLF in the range $3<z<5$ (see  Sect. \ref{sec:uvx} for a 
discussion on possible sources of this discrepancy).
The violet shaded area, which is our best estimate of the AGN ionizing emissivity, is the region 
included between the predictions of the $\log$N$_{\rm H}<21$ and $\log$N$_{\rm H}<22$ cm$^{-2}$ populations.
The shaded area is plotted in pink when the XLFs have been extrapolated ($z>$5).

Figure \ref{fig:emissivity} (right panel) shows the contribution to the emissivity 
produced by AGN at different $\rm N_H$: the $\log \rm N_H<$21 and $\log \rm N_H<$22 cm$^{-2}$ 
populations (same legend as the left panel) and the whole AGN population, 
including also Compton Thick sources 
\citep[][black dashed line]{Ueda2014},  
under the very extreme and unphysical assumption that all AGN 
(up to $\rm N_H=10^{26}$cm$^{-2}$) have $f_{esc}=1$.
Again, the emissivity has been drawn in grey when extrapolated at $z>5$.
The dotted-red horizontal line 
plotted in the right panel of Fig. \ref{fig:emissivity} shows, 
according to our analysis,
the UV ionizing AGN emissivity upper limit in the redshift range $5<z<6$,
obtained under the two assumptions 
that AGN with $\log \rm N_H<22$ cm$^{-2}$ contribute significantly to
reionization ($\langle f_{esc} \rangle$=1) and that the XLF remains constant for $z>5$.
In this case, the discrepancy between our upper limits 
and the results of \citet{Giallongo2015} is a factor of $\sim$4.
Even considering the contribution of the very absorbed AGN, 
a discrepancy of a factor $\sim$3 with the results of \citet{Giallongo2015} 
still remains. 

For reference, we also plot the evolution of the comoving emissivity as a function of redshift from 
\citet[][blue dashed line, their eq. 37; see also \citealt{Hopkins2007}]{HM2012}, \citet[][green dashed line, eq. 6 of their work]{KS2015} and \citet[][orange triple-dot-dashed line, eq. 1 of their work]{MH2015}.
We note that the agreement between our best estimate and other emissivities 
	in literature, which have been derived under different assumptions, is quite good.
As shown in Fig. \ref{fig:emissivity}, 
we found a high integrated local
AGN emissivity as recently proposed by \citet{MH2015}, a fact that can reduce the photon underproduction
crisis \footnote{The photon underproduction crisis is the finding of \citet{Kollmeier2014} of a five times higher \HI\ photoionization rate ($\Gamma_{\rm HI}$) at z = 0, obtained matching the observed properties of the low-redshift Ly$\alpha$ forest \citep[such as the Ly$\alpha$ flux decrement and the bivariate column density distribution of the Ly$\alpha$ forest; e.g.][]{Danforth2016}, than predicted by simulations which include state-of-the art models for the evolution of the UV background (UVB) \citep[e.g.][]{HM2012}. A similar investigation was carried out also by \citet{Shull2015}, who found a lower discrepancy (i.e. only a factor $\sim$ 2 higher) with the UVB model of \citet{HM2012}. 		} 
\citep[][see also \citealt{Shull2015}]{Kollmeier2014}. 
A precise analysis of this issue is however beyond the scope of this work.

Our best estimate is in agreement at $z<$2 with the $\epsilon_{912}$
proposed by \citet{MH2015}, while
the emissivities proposed by \citet{HM2012} and \citet{KS2015}
are in fair agreement at $2<z<6$ with our best estimate given the current uncertainties. 
In particular, the $\epsilon_{912}$ of the $\log$N$_{\rm H}<22$ cm$^{-2}$
agrees with the \citet{KS2015} estimate at $2<z<4.5$ while it is higher 
at $z>5$. The emissivity that we obtain considering only the $\log$N$_{\rm H}<21$ cm$^{-2}$ population, 
which represents our lower limit, is the lowest estimate at $2<z<5$ and agrees with the 
emissivity of \citet{HM2012} at $z >6$.
At variance, the QSO emissivity given by \citet{MH2015} is definitely larger than our estimate at 
$z>3$, as their analysis is based on the results by \citet{Giallongo2015}.

\section{Discussion}\label{sec:dis}

As discussed in the previous section, with our calculations we 
propose that
the range of possible values of the QSO ionizing emissivity
should lie in the shaded area highlighted in Fig. \ref{fig:emissivity}, i.e., in between the two limits obtained
considering the AGN population with $\log\rm N_H<21$ and $\log\rm N_H<22$ cm$^{-2}$(solid and dotted
black lines). 

We now compute the possible contribution of X--ray 
$\log\rm N_H<$21 and $\log\rm N_H<$22 cm$^{-2}$
AGN to the reionization of \HI\ residing in the IGM. 
The transition from a neutral to a fully ionized IGM is  
statistically described by a differential equation for the time 
evolution of the volume filling factor of the medium, $Q(z)$ \citep[see, e.g.][]{Madau1999}.
$Q(z)$ quantifies the level of the IGM porosity created 
by \HI\ ionization regions around radiative sources such as QSOs and SFGs.
The evolution of $Q(z)$ is given 
by the injection rate density of ionizing 
radiation minus the rate of radiative hydrogen recombination, 
whose temporal scale depends on the ionized hydrogen clumping factor
$C=\langle n^2_{HII} \rangle / \langle n_{HII}\rangle^2$.
Clumps that are thick enough to be self-shielded 
	from UV radiation do not contribute to 
	the recombination rate since they remain neutral.
A clumping factor of unity describes 
a homogeneous IGM. 
Results from recent hydrodynamical simulations, which take 
photo-heating of the IGM into account \citep{Pawlik2009,Raicevic2011},
show that C=3-10 are reasonable values for the clumping factor during the reionization.

\begin{figure}
	\centering
	\includegraphics[width=\columnwidth]{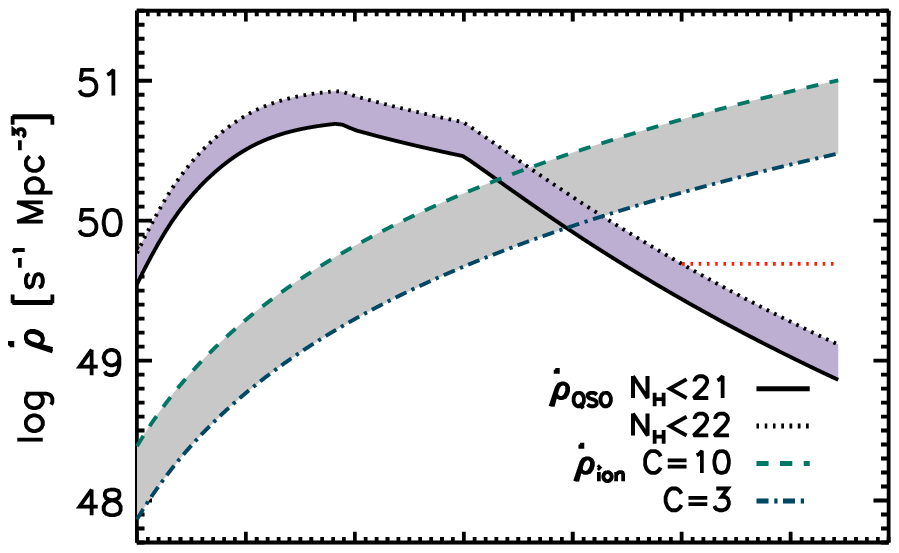}\vspace{-1.8cm}
	\includegraphics[width=\columnwidth]{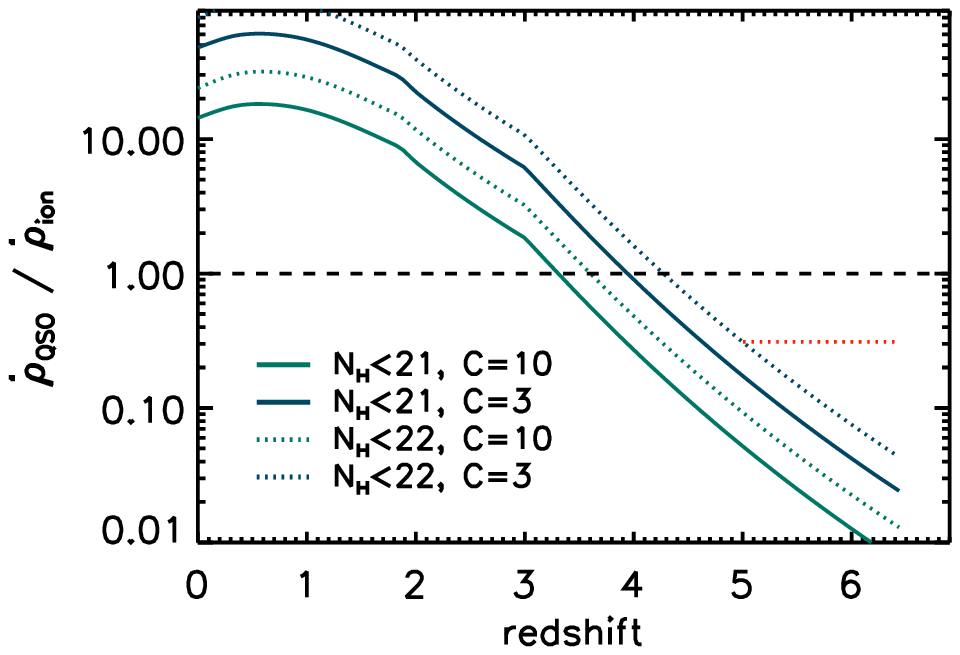}
	
	\caption{\textit{Top:} Comoving emission rate of hydrogen Lyman continuum
		photons from QSOs with $\log\rm N_H<21$ cm$^{-2}$ (black solid line) and with $\log\rm N_H<22$ cm$^{-2}$ (black dotted line), compared with the minimum rate needed to fully ionize the Universe with clumping factors C=3, 10 (blue dot-dashed, green dashed lines, respectively). \textit{Bottom:} Contribution 
		of QSO relative to the minimum rate computed with clumping factors C=3, 10 
		(blue and green lines, respectively). The dotted red line represents the maximal 
		AGN contribution (see text for details).}
	\label{fig:Nmin}
\end{figure}

By definition, the reionization finishes when all the 
hydrogen is fully ionized, i.e. when $Q=1$. 
Following \citet{Madau1999}, at any given epoch this condition  
translates into a critical value for the photon emission rate per unit cosmological comoving volume,
$\dot{\rho}_{ion}$, independently of the
(unknown) previous emission history of the Universe
\begin{equation}\label{eq:Nmin}
\dot{\rho}_{ion}(z)=10^{51.2} \left( \frac{C}{30}\right) \left(\frac{1+z}{6}\right)^3 \left( \frac{\Omega_b h^2_{70}}{0.0461} \right)^2 \, \rm Mpc^{-3} \, s^{-1}\, ,
\end{equation}
where we choose the normalization $\Omega_b h^2_{70}=0.0461$ from the results of the WMAP7 year data \citep{Komatsu2011}.
Only
rates above $\dot{\rho}_{ion}$ will provide enough UV photons to ionize the IGM by that epoch.

Fig. \ref{fig:Nmin} (top-panel) shows the results of this calculation 
assuming two different clumping factors, C=10 (green dashed line) and 3 (blue dot-dashed line). 
The grey shaded area 
indicates the possible range of $\dot{\rho}_{ion}$ obtained with
clumping factors between these two values.

We compare this minimum critical ionizing rate with those derived 
from the $\log \rm N_H<$21 and $\log\rm N_H<$22 cm$^{-2}$ AGN emissivities. We therefore compute
\begin{equation}
\dot{\rho}_{QSO}(z) = \int_{\nu_{HI}}^{\nu_{HeII}} \frac{\epsilon_\nu (z)}{\rm h \nu} \, d\nu \, ,
\end{equation} 
where $\rm h$ is the Planck's constant, $\nu_{HI}$ is the frequency at the 
Lyman limit (i.e. 1 ryd), $\nu_{HeII}=4\nu_{HI}$ and $\epsilon_\nu (z)$ is the 
QSO monochromatic ionizing emissivity 
which has been estimated from $\epsilon_{912}$ 
and then extrapolated between 1-4 ryd using the SED described in Sect. \ref{sec:uvx}.
Following \citet{ShankarMathur2007}, \citet{Fontanot2014},
	\citet{MH2015} and \citet{Cristiani2016}, the
upper limit on the integral is chosen at $4\nu_{HI}$ since 
more energetic photons are preferentially absorbed by helium atoms
\citep[see \citealt{Madau1999} for a complete discussion on the 
	advantages/limitations of this approximation, but see][for an alternative approach]{Grissom2014}.
The violet shaded area in the top panel of Fig. \ref{fig:Nmin} 
spans the possible values of $\dot{\rho}_{QSO}$ implied by 
our previous calculations on the emissivity.

The bottom panel in Fig. \ref{fig:Nmin} quantifies 
the contribution of the X--ray $\log\rm N_H<$21 (solid lines) and $\log\rm N_H<$22 cm$^{-2}$ (dotted lines)
populations relative to the minimum rate obtained from 
Eq. \ref{eq:Nmin} using C=3, 10 (blue and green curves, respectively).
At $z=6$, redshift of interest for the \HI\ reionization, we 
find that the contribution of ionizing AGN is little compared 
to the amount needed to fully ionize the IGM, 
with a maximun contribution of $\sim$7\% \citep[blue dotted curve, see also][for similar results]{ShankarMathur2007}.
If we consider the upper limit at $z>5$ on the QSO emissivity, shown in the right panel of Fig. \ref{fig:emissivity}, 
red dotted line (which has been derived
under the hypotheses that AGN with $\log\rm N_H<22$ cm$^{-2}$ contribute significantly to reionization
and that the XLF remains constant for $z>5$), then 
the contribution of ionizing AGN increases up to $\sim 30\%$ \citep[see also][]{ShankarMathur2007}.

We note that our best estimate of the AGN ionizing emissivity 
imply a dominant role of AGN only for $z\lesssim 4$ \citep[see also][and references therein]{Georgakakis2015,Cristiani2016}, slightly depending on the 
choice of the clumping factor C.
 
\section{Conclusions} \label{sec:conclusion}
The puzzling process of \HI\ reionization and the 
AGN contribution has been investigated
by using complete UV/optically-selected QSO
and X--ray selected AGN samples. 

In order to better constrain the faint end of 
the AGN LF at high redshift, we 
investigated whether the XLF could be used 
as an unbiased proxy of the ionizing AGN space density.
Indeed, X--ray selection offers a better control on the AGN faint end 
since it is less biased against obscuration. 
We employed the \citet{Ueda2014} XLF, which is 
computed in various absorption ranges, to derive a
matching between the UV/optical QLF and the X--ray
$\log\rm N_H\lesssim21 - 22$ cm$^{-2}$ AGN LF. 
The new \textit{Chandra} COSMOS Legacy Survey $z>3$ sample 
\citep{Marchesi2016b} is used to 
validate the extrapolation of \citet{Ueda2014} XLF 
beyond redshift 5, therefore enabling us to use the 
2--10 keV LF of \citet{Ueda2014} to compute the 
1 ryd comoving emissivity 
up to redshift $\sim$6. 
As expected in the traditional AGN unified model framework, 
when UV/optical data exist we found good agreement between 
the $\log\rm N_H\lesssim21-22$ cm$^{-2}$ XLF and the optically-selected QLF,
up to $z\sim4$. This matching implies that
the $\log\rm N_H\lesssim21-22$ cm$^{-2}$ XLF can be used as an unbiased 
proxy to estimate the density of ionizing AGN.

We found that 
the X--ray $\log\rm N_H<22$ cm$^{-2}$ LF at $z>4$
underpredicts by a factor $\sim1$ dex the 
the faint end of the UV LF 
derived using direct UV data. 
This discrepancy can be attributed to a contribution 
of UV emission from the AGN host galaxy 
whose amount is typical of galaxies 
	at break luminosity.

The use of the $\log\rm N_H\lesssim21-22$ cm$^{-2}$ XLFs allows us to 
measure the 1 ryd comoving QSO emissivity 
up to $z\sim 5$
without any luminosity extrapolation,
extending at $\sim$ 5 lower magnitudes than the limits probed by current UV/optical LFs.
The evolution of our proposed emissivity with redshift is 
in agreement also with the functional form proposed by 
\citet{MH2015} at $0<z<2$ and with
\citet{KS2015} and \citet{HM2012} at $2<z<6$ (all derived 
under different assumptions).
At variance, 
our estimate is smaller at $z>3$ than recently found by 
\citet{MH2015} who proposed an AGN-dominated scenario of 
\HI\ reionization. 
We found a high integrated local AGN emissivity as 
recently proposed by \citet{MH2015}.

Finally, we compare 
the photon emission rate necessary to ionize \HI\ with the critical 
value needed to keep 
the Universe ionized, independently of the
previous emission history of the Universe. 
Our findings are that 
the contribution of 
ionizing AGN at $z=6$ is little, $\sim 1 \% - 7 \%$, 
with a maximal contribution of $\sim$30\% under the unlikely assumption that the space density 
of $\log$N$\rm{_H}<$22 cm$^{-2}$ AGN remains constant at $z>5$. 
Our updated ionizing AGN emissivities thus exclude 
an AGN-dominated scenario at high redshifts, as instead recently suggested 
by other studies.

While this work makes use of the state of the art in terms of X--ray surveys, at the present day
it is impossible to extend this study at redshifts larger than 6. 
Even in the redshift range $5<z<6$
the available samples of X--ray selected high-redshift AGN still suffer 
of limited statistics, and are biased against relatively low luminosities 
($L_{2-10\,\rm{keV}}\leq$10$^{43}$ erg s$^{-1}$; below the LF break).
Only future facilities, like Athena \citep{Nandra2013} and the X--ray Surveyor
\citep{Vikhlinin2015}, will be able to collect sizable samples ($\sim$100s) of low luminosities ($L_{2-10\,\rm{keV}}<$10$^{43}$ ergs$^{-1}$) AGN at z$>$5 \citep{Civano2015}.

\section*{Acknowledgements}

We thank F. Fiore, F. Fontanot, E. Giallongo, A. Grazian and M. Volonteri for useful discussions.
We thank the anonymous referee for her/his valuable comments that improved 
	the quality of the manuscript.
We acknowledge funding from PRIN/MIUR-2010 award 2010NHBSBE and PRIN/INAF.
This work was partially supported by NASA Chandra grant number GO3-14150C and GO3-14150B 
(F.C., S.M.).





\bibliographystyle{mnras}
\bibliography{biblio} 







\bsp	
\label{lastpage}
\end{document}